\documentclass[12pt]{iopart}

\usepackage{geometry}                		
\usepackage{graphicx}				% Use pdf, png, jpg, or eps§ with pdflatex; use eps in DVI mode
\usepackage{amssymb}
\usepackage{iopams}
\usepackage{subfigure}
\usepackage{multirow}
\usepackage{cite}

\usepackage{footnote}
\usepackage[para]{footmisc}

\begin{document}

\title{New Limits on Double Electron Capture of $^{40}$Ca and $^{180}$W}

\author{
G~Angloher$^1$,
M~Bauer$^2$,
P~Bauer$^1$,
I~Bavykina$^1$,
A~Bento$^3$,
C~Bucci$^4$,
L~Canonica$^4$,
C~Ciemniak$^5$,
X~Defay$^5$,
G~Deuter$^2$,
A~Erb$^{5,6}$,
F~v.~Feilitzsch$^5$,
N~Ferreiro~Iachellini$^1$,
P~Gorla$^4$,
A~G\"{u}tlein$^7$,
D~Hauff$^1$,
P~Huff$^1$,
C~Isaila$^5$,
J~Jochum$^2$,
M~Kiefer$^1$,
M~Kimmerle$^2$,
H~Kluck$^7$,
H~Kraus$^8$,
J-C~Lanfranchi$^5$,
J~Loebell$^2$,
A~M\"unster$^5$,
C~Pagliarone$^4$,
F~Petricca$^1$,
S~Pfister$^5$,
W~Potzel$^5$,
F~Pr\"{o}bst$^1$,
F~Reindl$^1$,
S~Roth$^5$, 
K~Rottler$^2$, 
C~Sailer$^2$,
K~Sch\"{a}ffner$^9$,
J~Schieck$^7$,
J~Schmaler$^1$, 
S~Scholl$^2$,
S~Sch\"{o}nert$^5$,
W~Seidel$^1$,
M~v.~Sivers$^5$\footnote{Present address: Albert Einstein Center for Fundamental Physics, University of Bern, CH-3012 Bern, Switzerland},
L~Stodolsky$^1$,
C~Strandhagen$^2$,
R~Strauss$^1$,
A~Tanzke$^1$,
V~Tretyak$^{10,11}$,
H~H~Trinh~Thi$^5$,
C~T\"urko\v{g}lu$^7$,
M~Uffinger$^2$,
A~Ulrich$^5$,
I~Usherov$^2$,
S~Wawoczny$^5$,
M~Willers$^5$,
M~W\"{u}strich$^1$,
A~Z\"{o}ller$^5$
}

\address{$^1$ Max-Planck-Institut f\"ur Physik, D-80805 M\"unchen, Germany}
\address{$^2$ Eberhard-Karls-Universit\"{a}t T\"{u}bingen, D-72076 T\"{u}bingen, Germany}
\address{$^3$ Departamento de Fisica, Universidade de Coimbra, P3004 516 Coimbra, Portugal}
\address{$^4$ INFN, Laboratori Nazionali del Gran Sasso, I-67010 Assergi, Italy}
\address{$^5$ Physik-Department and Excellence Cluster Universe, Technische Universit\"{a}t M\"{u}nchen, D-85747 Garching, Germany}
\address{$^6$ Walther-Mei{\ss}ner-Institute for Low Temperature Research, 85748 Garching, Germany}
\address{$^7$ Institut f\"ur Hochenergiephysik der \"Osterreichischen Akademie der Wissenschaften, A-1050 Wien, Austria and Atominstitut, Vienna University of Technology, A-1020 Wien, Austria}
\address{$^8$ Department of Physics, University of Oxford, Oxford OX1 3RH, United Kingdom}
\address{$^9$ GSSI - Gran Sasso Science Institute, I-67100 L'Aquila - Italy}
\address{$^{10}$ Institute for Nuclear Research, MSP 03680 Kyiv, Ukraine}
\address{$^{11}$ INFN, sezione di Roma, I-00185 Rome, Italy}

\ead{moritz.vonsivers@lhep.unibe.ch}

\begin{abstract}
We analyzed low-background data from the CRESST-II experiment with a total net exposure of 730\,kg\,days to extract limits on double electron capture processes. We established new limits for $^{40}$Ca with $T_{1/2}^{2\nu2K}>9.9\times10^{21}$\,y and $T_{1/2}^{0\nu2EC}>1.4\times10^{22}$\,y and for $^{180}$W with $T_{1/2}^{2\nu2K}>3.1\times10^{19}$\,y and $T_{1/2}^{0\nu2EC}>9.4\times10^{18}$\,y at 90\% CL. Depending on the process, these values improve the currently best limits by a factor of $\sim1.4\mbox{--}30$.

\end{abstract}

%\pacs{1315, 9440T}
%\keywords{magnetic moment, solar neutrinos, astrophysics}
\submitto{\jpg}

\maketitle

\section{Introduction}
Double electron capture (2EC) is a rare nuclear decay where a nucleus $({\rm A, Z})$  captures two electrons from the inner atomic shells thereby lowering its charge by two units transforming into $({\rm A, Z}-2)^{**}$. The two stars denote the excitation of the atomic shell due to the electron vacancies and a possible excitation of the nucleus.
In principle, there are two modes for the decay, two neutrino double electron capture (2$\nu$2EC) as shown in (\ref{eq:2nu2EC}) and zero neutrino double electron capture (0$\nu$2EC) presented in (\ref{eq:0nu2EC}):
\begin{eqnarray}
({\rm A, Z}) + 2{\rm e}^{-} \rightarrow ({\rm A, Z}-2)^{**} + 2\nu_{\rm e}, \label{eq:2nu2EC}\\
({\rm A, Z}) + 2{\rm e}^{-} \rightarrow ({\rm A, Z}-2)^{**}.\label{eq:0nu2EC}
\end{eqnarray}
So far 2$\nu$2EC has only been observed for $^{130}$Ba in geochemical experiments \cite{meshik01,pujol09}. In addition, there is a $2.5\sigma$ evidence for the process in $^{78}$Kr from a low-background proportional counter \cite{gavrilyuk13}.
Process (\ref{eq:0nu2EC}) is forbidden in the Standard Model of particle physics, as it violates the lepton number conservation by two units. 
Similar to neutrinoless double beta decay (0$\nu$2$\beta$), the observation of 0$\nu$2EC would prove the Majorana character of the neutrino \cite{sujkowski03}. Limits on 0$\nu$2EC or 0$\nu$2$\beta$ can be used to constrain the effective neutrino mass $m_{\beta\beta}$ and investigate the neutrino mass hierarchy. The experimental search for lepton number violating processes is mainly focused on 0$\nu$2$\beta$ where the predicted half-life is more favorable because of phase space arguments. In general, the initial and final states in (\ref{eq:0nu2EC}) will have different masses. Therefore, energy conservation requires an additional photon to be emitted which leads to very large predicted half-lives.
However, in case of a mass degeneracy between the initial and final state there is a resonant enhancement of the decay rate. This can make the process competitive to searches for 0$\nu$2$\beta$ \cite{sujkowski03,krivoruchenko10}. In the recent past resonantly enhanced 0$\nu$2EC has been the topic of many theoretical \cite{krivoruchenko10,fang12,suhonen12,rodriguez12,maalampi13} and experimental \cite{barabash06,barabash08,barabash09,belli11,belli13,belli14} studies.

In this paper, we derive experimental limits on the half-lives of 2$\nu$2EC and 0$\nu$2EC processes for $^{40}$Ca and $^{180}$W. The latter is one of the best candidates to observe resonant 0$\nu$2EC \cite{krivoruchenko10,kotila14}. A summary of the processes studied in this work is shown in table~\ref{tab:processes}. For the 2$\nu$2EC transition to the ground state the atom de-excites via the emission of X-rays and/or Auger electrons, and the observable energy equals the sum of the binding energies of the captured electrons. Because K electrons are closest to the nucleus, the most probable process is double K-capture (2$\nu$2K)\footnote{Using the code CAPTURAT \cite{capturat}, the probability of 2K (2L) capture can be estimated as 0.85 (0.01) for $^{40}$Ca and 0.40 (0.14) for $^{180}$W.}, hence the observed energy equals 2E$_K$. For 0$\nu$2EC the total observable energy is always given by the Q-value of the decay. Table~\ref{tab:processes} also summarizes the currently best experimental limits on the half-life along with some theoretical predictions.

\fulltable{\label{tab:processes}Double electron capture processes studied in this work. The last two columns show respectively the currently best experimental limits on the half-life along with theoretical predictions.} 
\br     
 & & & Observable & & \\
 \centre{1}{Isotope} & \centre{1}{Abundance (\%)} & \centre{1}{Process} & \centre{1}{Energy (keV)} & \centre{1}{T$_{1/2}^{\rm exp}$ (y) (90\% CL)} & \centre{1}{T$_{1/2}^{\rm th}$ (y)} \\

\mr
 \multirow{2}{*}{$^{40}$Ca} & \multirow{2}{*}{96.94(16) \cite{berglund11}} & 0$\nu$2EC & 193.51(2) \cite{wang12} & $>3.0\times10^{21}$ \cite{belli99} & - \\ 
& & 2$\nu$2K & 6.4 \cite{xrdbl} &  $>7.3\times10^{21}$ \cite{belli99}$^{\rm a}$ &  $1.2\times10^{33}$ \cite{ching84}\\
  \multirow{2}{*}{$^{180}$W} &  \multirow{2}{*}{0.12(1) \cite{berglund11}}& 0$\nu$2EC & 143.27(20) \cite{droese11} & $>1.3\times10^{18}$ \cite{belli11} & $(1.3-1.8)\times10^{31}$ \cite{fang12}$^{\rm b}$  \\
& & 2$\nu$2K & 130.7 \cite{xrdbl} &  $>1.0\times10^{18}$ \cite{belli11} & $\sim2.5\times10^{28}$ \cite{iachello}\\
\br
\end{tabular*}
\noindent $^{\rm a}$ The limit in \cite{belli99} is given for 2$\nu$2EC assuming a probability of 0.81 for double K-capture. \\
\noindent $^{\rm b}$ The predicted half-life in \cite{fang12} is calculated for $m_{\beta\beta}=50$\,meV.
%\endfulltable
%\end{indented}
\end{table}

%\begin{table}[htdp]
%\begin{center}
%\begin{tabular}{cccccc}
%\hline
%\textbf{Isotope} & \textbf{Abundance (\%)} & \textbf{Process} & \textbf{Energy (keV)} & \textbf{T$_{1/2}$ (exp.)} & \textbf{T$_{1/2}$ (pred.)} \\
%\hline
%\multirow{2}{*}{$^{40}$Ca} & \multirow{2}{*}{96.9} & 0$\nu$2EC & 193.6 & $>3.0\times10^{21}$ (90\% CL) \cite{belli99} & - \\ 
%  & & 2$\nu$2K & 6.4 &  $>7.3\times10^{21}$ (90\% CL)\footnote{The limit in \cite{belli99} is given for 2$\nu$2EC assuming a probability of double K-capture of 0.81.}
% \cite{belli99} &  $1.2\times10^{33}$ \cite{ching84}\\
% \multirow{2}{*}{$^{180}$W} & \multirow{2}{*}{0.12}& 0$\nu$2EC & 144 & $>1.3\times10^{18}$ (90\% CL)\cite{belli11} & $3.0\times10^{22}-3.8\times10^{32}$ \cite{krivoruchenko10,rodriguez12}  \\
%& & 2$\nu$2K & 130.7 &  $>1.0\times10^{18}$ (90\% CL)\cite{belli11} & -\\
%\hline
%\end{tabular}
%\end{center}
%\label{tab:processes}
%\caption{Double electron capture (2EC) processes studied in this work.}
%
%\end{table}%

\section{Experiment \& Data Analysis}
CRESST-II (Cryogenic Rare Event Search with Superconducting Thermometers) \cite{angloher15} aims at the direct detection of dark matter. The detector consists of scintillating bolometers based on CaWO$_4$ crystals. A detailed description of the setup can be found elsewhere \cite{angloher09}. Between 2009 and 2011, a total net exposure of 730\,kg\,days has been collected with eight detector modules. The data were previously analyzed for a possible WIMP signal in the form of low-energy nuclear recoils \cite{angloher12}. Here we use these data to derive limits on the double electron capture of $^{40}$Ca and $^{180}$W. 

Basic data quality cuts were applied to the data set as described in \cite{angloher12}. In addition, only single-scatter events, i.e. events with no coincident signal in any other detector module or the muon veto were accepted. The energy range extends from the trigger threshold (around 4\,keV) to 300\,keV. The latter was set as an upper limit for the WIMP analysis where signal events are only expected below 40\,keV. 

The energy calibration of the detectors was performed with 122\,keV $\gamma$-rays from a $^{57}$Co calibration source. The calibration was extended to lower energies with the help of heater pulses which were injected to the detector \cite{angloher09}. After this calibration some deviations in the position of known $\gamma$-lines in the background spectra were found at energies $\gtrsim$150\,keV. Therefore, the spectra were re-calibrated by fitting the position of these $\gamma$-lines with a second order polynomial function. After re-calibration, deviations of the observed $\gamma$-lines from the literature values \cite{nudat} were $\lesssim0.5$\,keV. 
\begin{figure}[htbp]
\begin{center}
\subfigure[]{\includegraphics[width=0.49\textwidth]{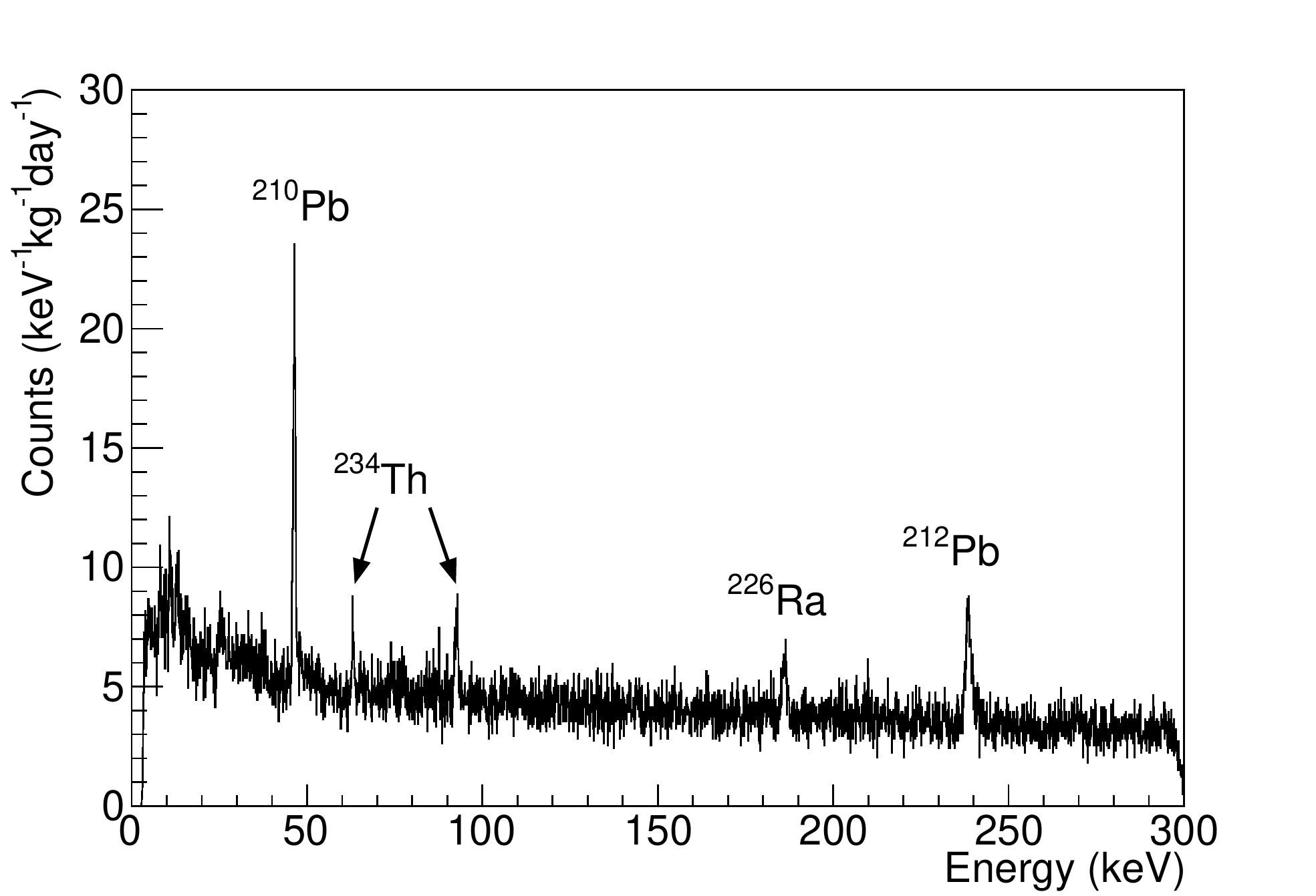}}
\subfigure[]{\includegraphics[width=0.49\textwidth]{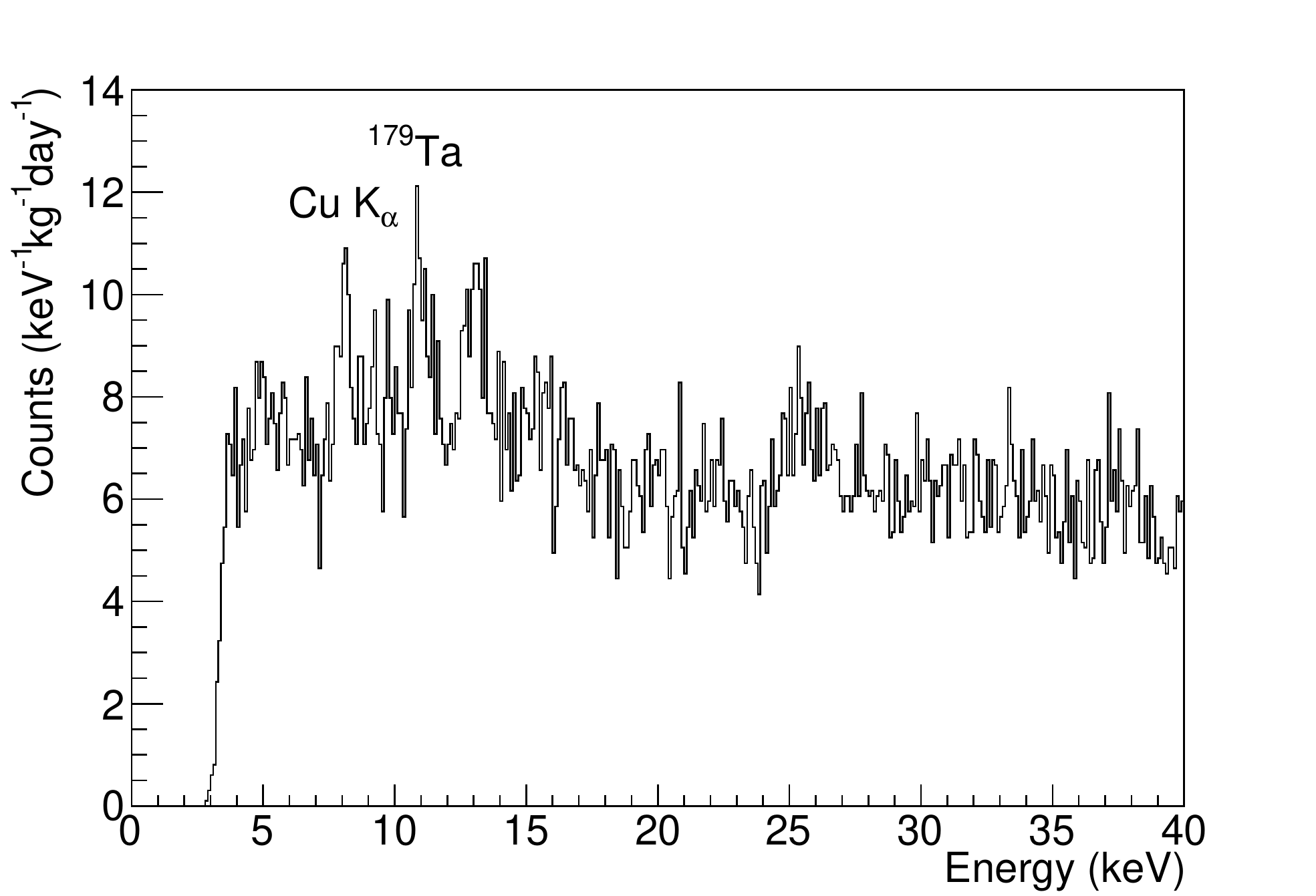}}
\caption{Background spectrum of the detector \textit{Ch47}. The visible $\gamma$-lines originate from external radioactivity and cosmogenic activation. Panel (b) shows a zoom to the low-energy region.}
\label{fig:spectrum}
\end{center}
\end{figure}

Figure \ref{fig:spectrum}(a) shows a typical spectrum of a single detector module. The $\gamma$-lines are due to external radioactivity from $^{212}$Pb (238.6\,keV \cite{nudat}), $^{226}$Ra (186.2\,keV \cite{nudat}) and $^{210}$Pb (46.5\,keV \cite{nudat}). In the low energy region (see figure~\ref{fig:spectrum}(b)) weak lines from Cu fluorescence (8.0\,keV \cite{nudat}) and the L-capture of $^{179}$Ta (11.3\,keV \cite{xrdbl}) are visible. The latter stems from cosmogenic activation of the CaWO$_4$ crystals \cite{strauss14}. In addition, a so far unidentified line at $\sim13$\,keV can be seen.

The energy resolution of each detector was modeled individually by the following equation:
\begin{equation}
\sigma(E)=\sqrt{{\sigma_0}^2+{\sigma_1}^2 E + {\sigma_2}^2 E^2},
\label{eq:resolution}
\end{equation}
where $\sigma_0$ represents energy-independent contributions which influence the baseline noise, the $\sigma_1$ term reflects Poisson-like contributions and $\sigma_2$ stands for higher-order contributions (e.g. position dependence).
Here the parameter $\sigma_0$ is derived from the resolution of the lowest injected heater pulses. The other parameters are obtained by fitting (\ref{eq:resolution}) to the resolution of all $\gamma$-lines in the background spectra. Typically the 1-$\sigma$ energy resolution at 122\,keV is 0.52\,keV.

For all studied processes, there is a high probability that the released X-rays (Auger electrons) and/or $\gamma$-rays (conversion electrons) will be fully absorbed inside the detectors, hence the expected signal is a peak at the energy given in table~\ref{tab:processes}. The detection efficiency $\epsilon$ for all processes was obtained by a Geant4 \cite{agostinelli02} simulation which simulates the energy deposition in a cylindrical 300\,g CaWO$_4$ crystal of 40\,mm height and 40\,mm diameter. The initial kinematics of events were taken from the DECAY0 event generator \cite{ponkratenko00}. Table~\ref{tab:efficiency} summarizes the results of the efficiency simulation.
\Table{\label{tab:efficiency}Detection efficiency $\epsilon$ for the full energy absorption peak obtained from a Geant4 simulation. The quoted uncertainties are purely statistical.}
\br
Isotope & Process & \centre{1}{Detection efficiency $\epsilon$} \\
\mr
\multirow{2}{*}{$^{40}$Ca} & 0$\nu$2EC & $0.877\pm0.001$ \\ 
  & 2$\nu$2K & $1.0\pm0.001$ \\
 \multirow{2}{*}{$^{180}$W} & 0$\nu$2EC & $0.938\pm0.001$  \\
& 2$\nu$2K & $0.938\pm0.001$ \\
\br
\end{tabular}
\end{indented}
\end{table}
A Bayesian approach was chosen for the analysis using the Bayesian Analysis Toolkit (BAT) \cite{caldwell08}. 
The spectra were fitted with a "signal$+$background" model $M$ in an energy range $\pm5\sigma$ around the expected signal peak. Signal and background were modeled with a Gauss function and a constant term, respectively:
\begin{equation}
M=\frac{\Gamma \eta \epsilon N_{\rm A} t}{M_{\rm CaWO_4}\sqrt{2\pi}\sigma_{\rm sig}}\rme^{-\frac{(x-\mu_{\rm sig})^2}{2\sigma_{\rm sig}^2}}+c_{\rm bkg}.
\label{eq:M}
\end{equation}
Here $\Gamma$ is the decay rate, $\epsilon$ is the detection efficiency for the full energy peak, N$_{\rm A}$ is the Avogadro number, $\eta$ is the natural abundance of the isotope, $t$ is the exposure (in kg\,days) and m$_{{\rm CaWO}_4}$ is the molar mass of CaWO$_4$.
In three detectors (\textit{Ch29}, \textit{Ch33} and \textit{Ch43}), due to their worse resolution, the 186.2\,keV peak from $^{226}$Ra lies in the $\pm5\sigma$ fit range of the peak from 0$\nu$2EC of $^{40}$Ca expected at 193.6\,keV. In these cases, another Gauss function was included in the model to account for the 186.2\,keV peak:
\begin{equation}
\fl M_{0\nu2{\rm EC}, ^{40}{\rm Ca}}^{Ch29,Ch33,Ch43}=\frac{\Gamma \eta \epsilon N_{\rm A} t}{M_{\rm CaWO_4}\sqrt{2\pi}\sigma_{ \rm sig}}\rme^{-\frac{(x-\mu_{\rm sig})^2}{2\sigma_{\rm sig}^2}}+c_{\rm bkg}+\frac{a_{\rm bkg}}{\sqrt{2\pi}\sigma_{\rm bkg}}\rme^{-\frac{(x-\mu_{\rm bkg})^2}{2\sigma_{\rm bkg}^2}}.
\label{eq:M_0nu2K}
\end{equation}
For three detectors (\textit{Ch05}, \textit{Ch29} and \textit{Ch43}) the background in the low-energy region around the expected peak of 2$\nu$2K of $^{40}$Ca at 6.4\,keV is not well described by a simple constant, i.e. the fit returns a very small p-value numerically compatible with zero. In these cases a more conservative approach was chosen to calculate an upper limit on the half-life. The spectrum was fitted in the energy range $\pm1\sigma$ around the expected signal using only a Gaussian for the signal without making any assumptions on the background:
\begin{equation}
M_{2\nu2{\rm K}, ^{40}{\rm Ca}}^{Ch05,Ch29,Ch43}=\frac{\Gamma \eta \epsilon N_{\rm A} t}{M_{\rm CaWO_4}\sqrt{2\pi}\sigma_{\rm sig}}\rme^{-\frac{(x-\mu_{\rm sig})^2}{2\sigma_{\rm sig}^2}}.
\label{eq:M_2nu2K}
\end{equation}
The best fit values for the parameters $\vec{\lambda}$ were obtained by maximizing the total posterior probability distribution function (\textit{pdf}):
\begin{equation}
P(\vec{\lambda} \mid \vec{D}) = \frac{P(\vec{D}\mid \vec{\lambda}) P_{\rm 0}(\vec{\lambda})}{\int P(\vec{D} \mid \vec{\lambda}) P_0(\vec{\lambda})\,\rmd\vec{\lambda}},
\end{equation}
where $\vec{\lambda}$ are the model parameters and $\vec{D}$ are the data.
The likelihood $P(\vec{D} \mid \vec{\lambda})$ is calculated assuming Poissonian uncertainties on the expectation value in each bin.
$P_0(\vec{\lambda})$ are the prior probabilities of the parameters. Uniform priors were used for the decay rate $\Gamma$, the number of background counts $a_{\rm bkg}$ and the constant $c_{\rm bkg}$. To include systematic uncertainties of the peak positions, energy resolution and natural abundances, Gaussian priors were chosen for the parameters $\mu_{\rm sig}$, $\mu_{\rm bkg}$, $\sigma_{\rm sig}$, $\sigma_{\rm bkg}$ and $\eta$. For the means of the signal and background peaks, $\mu_{\rm sig}$ and $\mu_{\rm bkg}$, the prior was chosen according to the uncertainty of the energy calibration which was derived from the confidence band of the fit function to the energy calibration. For the parameter $\mu_{\rm sig}$ also the uncertainty of the Q-value of the 0$\nu$2EC process (see table~\ref{tab:processes}) was included. In the case of 2$\nu$2K the additional uncertainties of the electron binding energies are negligible.
The priors of the standard deviations $\sigma_{\rm sig}$ and $\sigma_{\rm bkg}$ were determined from the fit function and corresponding confidence band of the energy resolution.
For the natural abundance $\eta$ we took the uncertainty as listed in table~\ref{tab:processes}.
All parameters in (\ref{eq:M})-(\ref{eq:M_2nu2K}) were constrained to physically allowed positive values. 

The analysis was carried out individually for each detector module. In addition, a combined fit to several detectors was performed. 
In the fit model the decay rate $\Gamma$ was a common parameter to all detectors. To obtain the posterior \textit{pdf} of the combined fit the likelihoods were multiplied for all $N$ detector modules:
\begin{eqnarray}
P(\vec \lambda_{\rm tot} \mid \vec D_{\rm tot}) =  \frac{P(\vec D_{\rm tot} \mid \vec \lambda_{\rm tot}) P_{\rm 0}(\vec \lambda_{\rm tot})}{\int P(\vec D_{\rm tot} \mid \vec \lambda_{\rm tot} ) P_{\rm 0}(\vec \lambda_{\rm tot})\,\rmd\vec \lambda_{\rm tot}}, \\
P( \vec D_{\rm tot} \mid \vec \lambda_{\rm tot}) =  \prod\limits_{i=1}^{N}P(\vec D_i \mid \vec \lambda_{\rm tot}).
\end{eqnarray}
The estimated experimental sensitivity of all detectors is $\sim10^{21}$\,y and $\sim10^{18}$\,y for the half-lives of $^{40}$Ca and $^{180}$W, respectively. These values are several orders of magnitude lower than the theoretical predictions of the half-lives (see table~\ref{tab:processes}) leaving no chance for the possible observation of a signal. Lower limits on the half-lives were calculated from the posterior \textit{pdf} of the decay rate $\Gamma$:
\begin{equation}
P(\Gamma \mid D) = \int P (\vec{\lambda} \mid D)\,\rmd\vec{\lambda}\vert_{\lambda_i\neq \Gamma}.
\end{equation}
The 90\% CL upper limit $\Gamma_{\rm lim}$ on the decay rate was calculated by:
\begin{equation}
0.9=\int\limits_{0}^{\Gamma_{\rm lim}} P(\Gamma \mid D)\,\rmd \Gamma.
\end{equation}
The limit on the half-life T$_{1/2}$ was then calculated according to the following equation:
\begin{equation}
T_{1/2}>\frac{ln(2)} {\Gamma_{\rm lim}}.
\end{equation}

\section{Results \& Discussion}
\begin{figure}[b!]
\begin{center}
\subfigure[]{\includegraphics[width=0.49\textwidth]{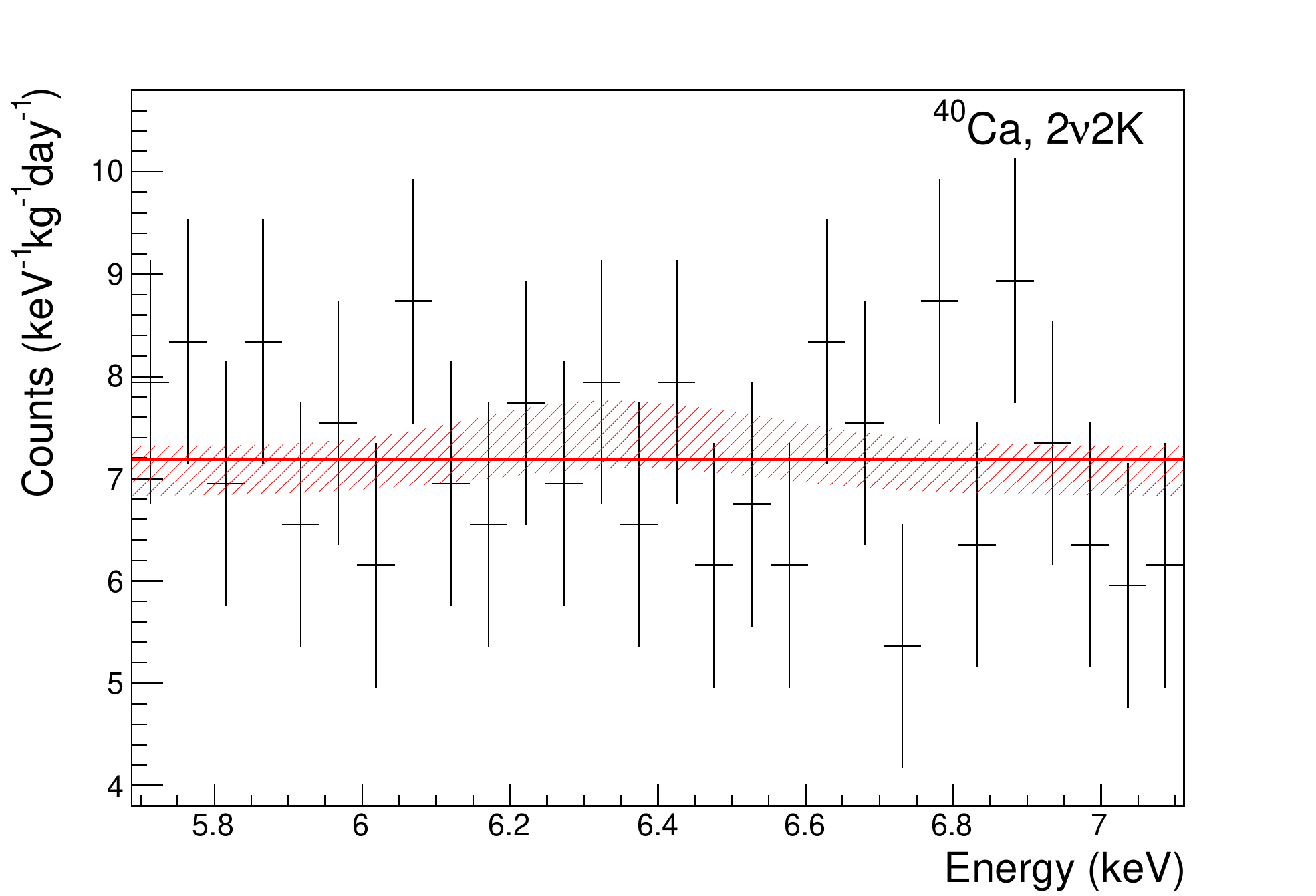}}
\subfigure[]{\includegraphics[width=0.49\textwidth]{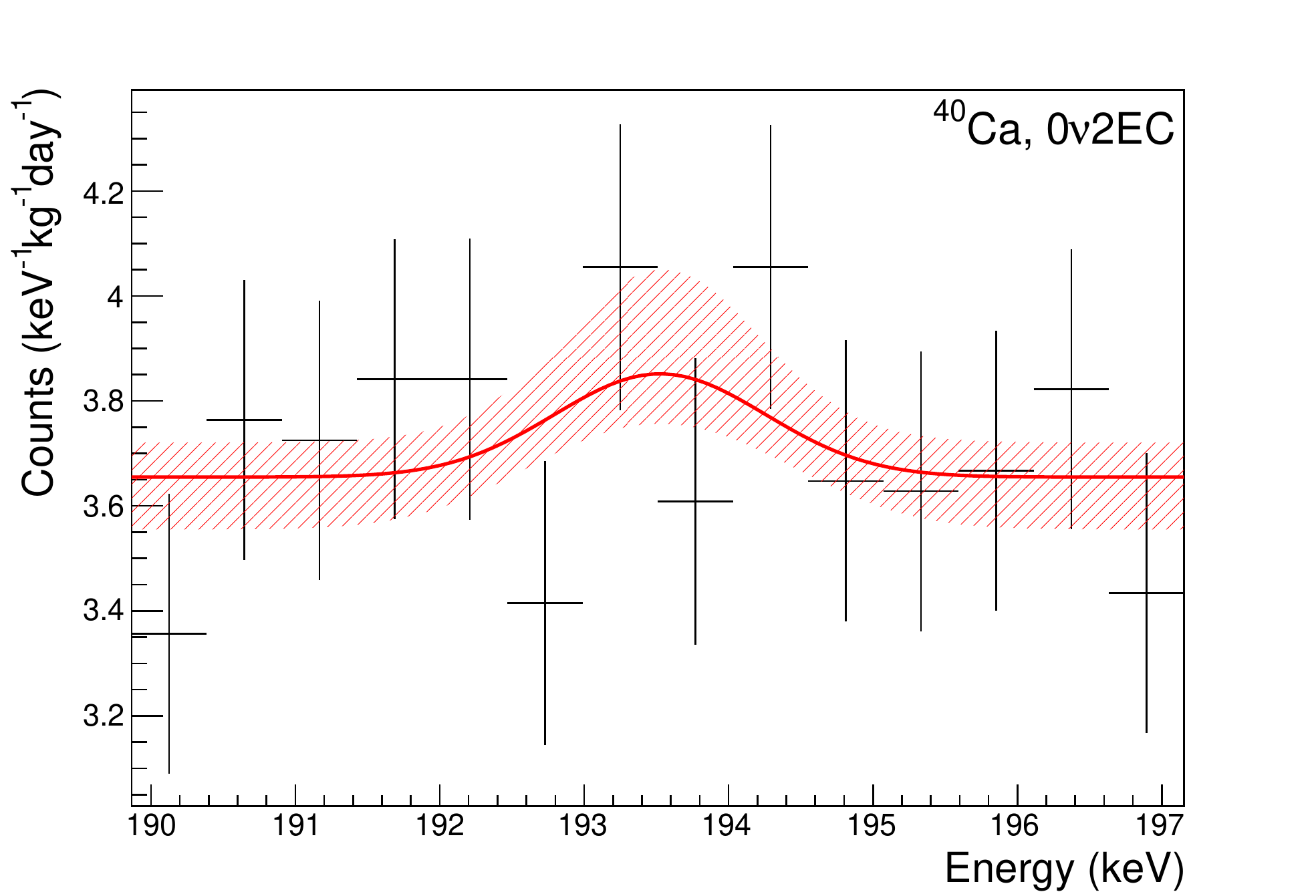}}
\subfigure[]{\includegraphics[width=0.49\textwidth]{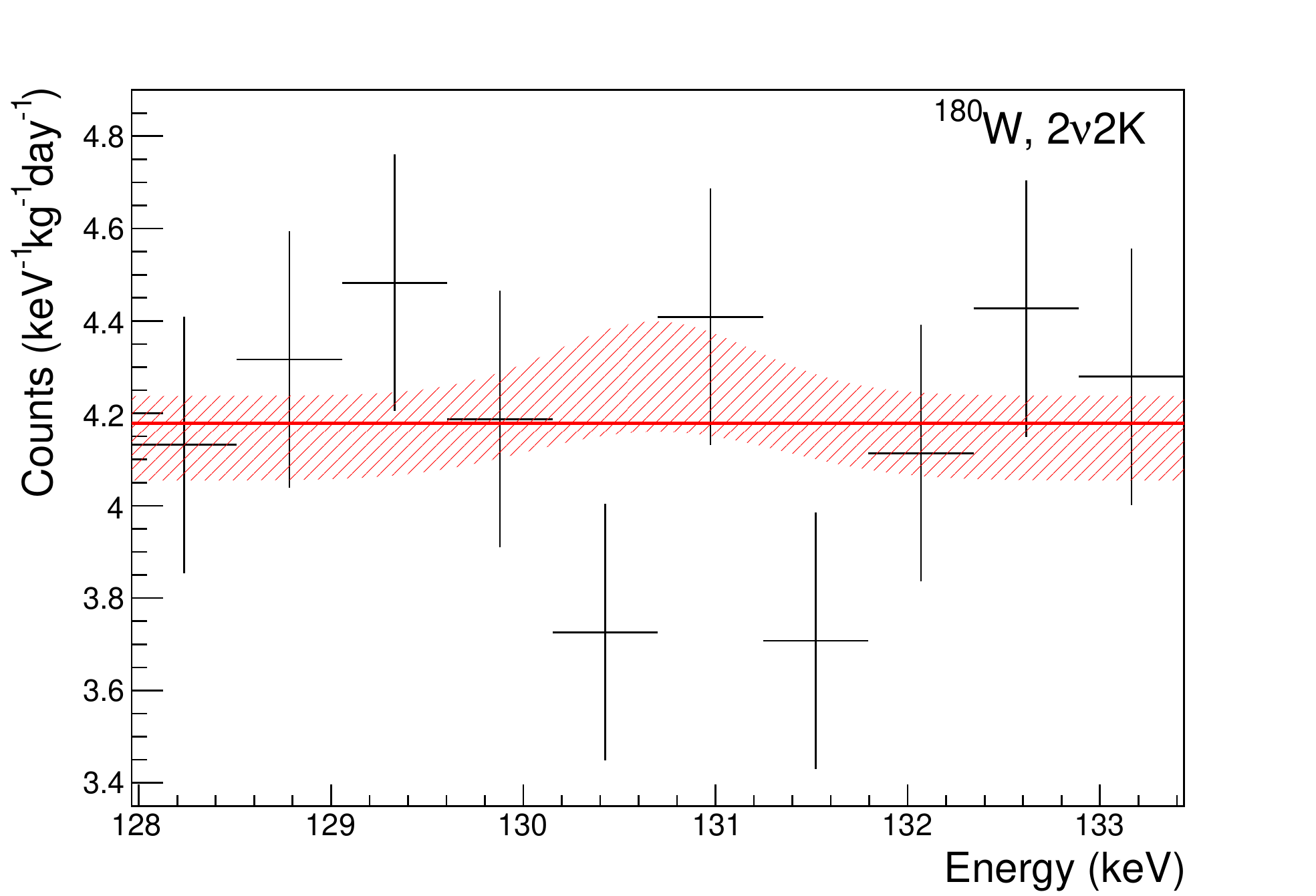}}
\subfigure[]{\includegraphics[width=0.49\textwidth]{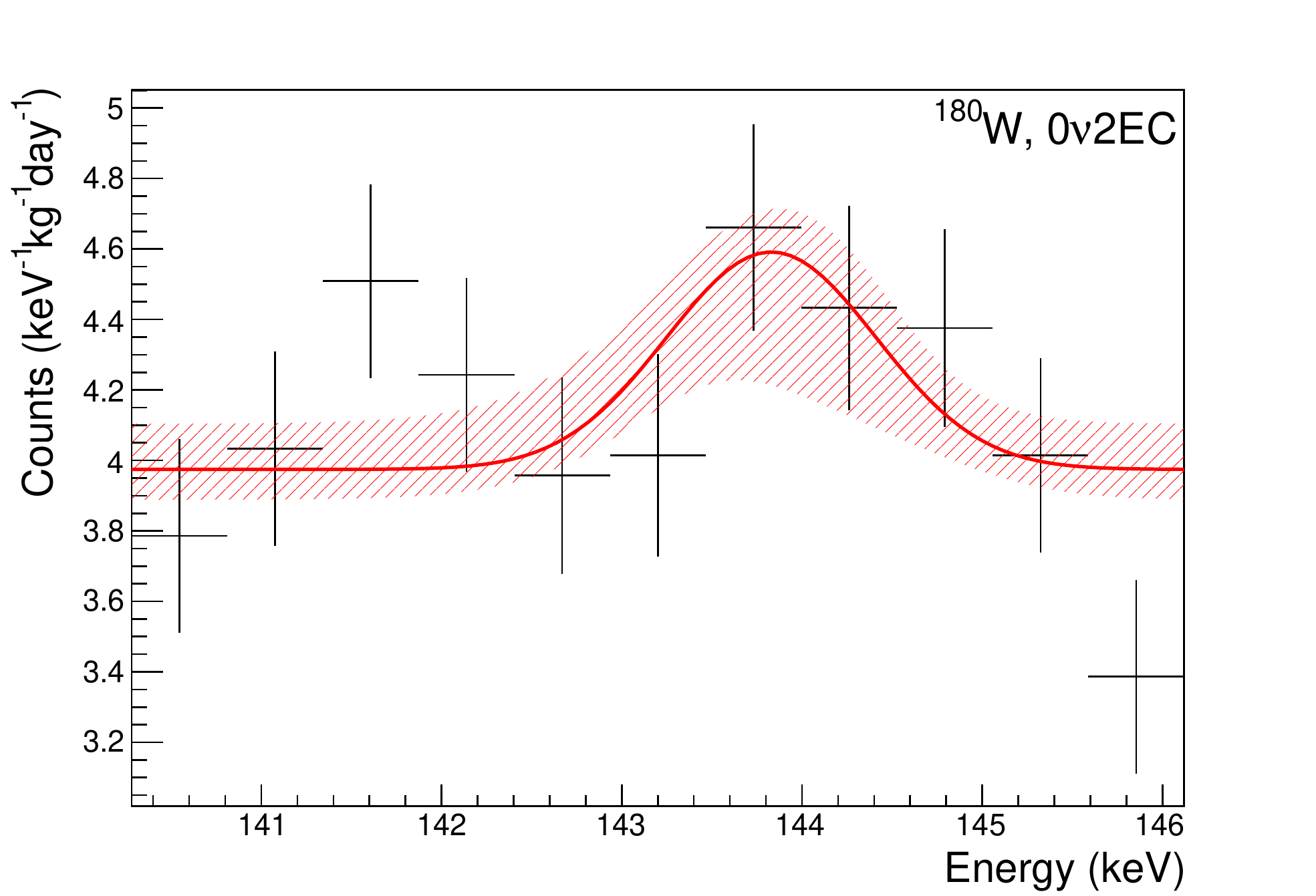}}
\caption{Best fit of the signal$+$background model $M$ to the spectrum of detector \textit{Ch47}. The hatched area indicates the 68\% uncertainty band. }
\label{fig:best_fit}
\end{center}
\end{figure}

Figure~\ref{fig:best_fit} shows the best fit of all studied processes for a single detector module. The results of all detectors are summarized in tables~\ref{tab:results_Ca40} and \ref{tab:results_W180}. The goodness-of-fit was evaluated by calculating the p-value as described in \cite{beaujean11} and is shown in brackets in tables~\ref{tab:results_Ca40} and \ref{tab:results_W180}. In most cases the p-value ranges between 0.5-0.9 showing that the data are well described by the fit model. In the combined fit for 2$\nu$2K of $^{40}$Ca we excluded the detector modules \textit{Ch05}, \textit{Ch29} and \textit{Ch43} where, as explained above, the background is not well modelled by a constant.
For 2$\nu$2K of $^{40}$Ca the strongest limit on the half-life is $>9.92\times10^{21}$\,y. 
This value improves the currently best limit only marginally. In case of 0$\nu$2EC of $^{40}$Ca the new half-life limit $>1.40\times10^{22}$\,y improves the currently best limit by a factor of $\sim5$. For 2$\nu$2K of $^{180}$W the new limit $>3.13\times10^{19}$\,y is leading to a large improvement by a factor of $\sim30$. The half-life limit $>9.39\times10^{18}$\,y for 0$\nu$2EC of $^{180}$W improves the previous limit by a factor of $\sim7$.

\Table{\label{tab:results_Ca40}Extracted limits on the half-life $T_{1/2}$ for 2EC of $^{40}$Ca. The values in brackets show the p-value of the corresponding fit. The analysis was performed individually for all detector modules and for a combination of several detectors. For details see text.}
\br
& \centre{2}{90\% CL Limit on T$_{1/2}$ ($10^{21}$\,y)}  \\
Detector & \centre{1}{2$\nu$2K} & \centre{1}{0$\nu$2EC} \\
\mr
Ch05  &  0.38 (0.024) & 4.59 (0.623) \\ 
Ch20  & 1.76 (0.830) & 2.79 (0.611) \\
Ch29  &  0.27 (0.324) & 3.40 (0.150) \\ 
Ch33  &  3.40 (0.466) & 4.55 (0.978) \\ 
Ch43  & 0.10 (0.001) & 3.20 (0.984) \\ 
Ch45  & 5.19 (0.242) & 5.14 (0.861) \\ 
Ch47  & 9.92 (0.919) & 3.54 (0.905) \\ 
Ch51  & 0.71 (0.714) & 5.63 (0.780) \\ 
\hline
Combined Fit  & 7.96 (0.022) & 14.0 (0.930) \\ %0.285 (0.011) &  () \\ 
\br
\end{tabular}
\end{indented}
\end{table}
\Table{\label{tab:results_W180}Extracted limits on the half-life $T_{1/2}$ for 2EC of $^{180}$W. Other details as in table~\ref{tab:results_Ca40}.}
\br
& \centre{2}{90\% CL Limit on T$_{1/2}$ ($10^{18}$\,y)}  \\
Detector & \centre{1}{2$\nu$2K} & \centre{1}{0$\nu$2EC} \\
\mr
Ch05  & 4.39 (0.646) & 9.39 (0.785)  \\ 
Ch20  & 5.96 (0.908) & 4.68 (0.520) \\
Ch29  & 4.46 (0.710) & 1.78 (0.067)  \\ 
Ch33  & 5.57 (0.909) & 4.66 (0.545)  \\ 
Ch43  & 4.19 (0.813) & 3.77 (0.756)  \\ 
Ch45  & 13.0 (0.558) & 3.61 (0.758)  \\ 
Ch47  & 10.3 (0.513) & 3.27 (0.401)  \\ 
Ch51  & 5.68 (0.583) & 3.42 (0.844)  \\ 
\hline
Combined Fit  &  31.3 (0.902) & 8.08 (0.734) \\ %8.93 (0.949) \\ 
\br
\end{tabular}
\end{indented}
\end{table}

\section{Summary \& Conclusion}
Using low-background data from the CRESST-II experiment we have extracted new limits on the half-life of 2$\nu$2K and 0$\nu$2EC for $^{40}$Ca and $^{180}$W. Depending on the process, the new values improve the currently best limits by a factor of $\sim1.4\mbox{--}30$. Although the limits are still far from theoretical predictions this result highlights the feasibility to study double beta processes with CRESST-II detectors. Further improvement on the half-life limits can be expected from the data taken with new CRESST detectors with improved radiopurity \cite{munster14,strauss14}. In addition, an analysis of the high energy region to study the double beta decays of $^{46}$Ca, $^{48}$Ca and $^{186}$W is planned.

\ack
We are grateful to LNGS for their generous support of CRESST, in particular to Marco Guetti for his constant assistance. We gratefully acknowledge F. Iachello for providing us with the calculation of $T_{1/2}^{2\nu2K}$ of $^{180}$W prior to publication.
This research was supported by the DFG cluster of excellence 'Origin and Structure of the Universe', the Helmholtz Alliance for Astroparticle Physics, the Maier-Leibnitz-Laboratorium (Garching) and by the BMBF: Project 05A11WOC EURECA-XENON.

\section*{References}
\bibliographystyle{iopart-num.bst}
\bibliography{Bibliography.bib}

\providecommand{\newblock}{}
\begin{thebibliography}{10}
\expandafter\ifx\csname url\endcsname\relax
  \def\url#1{{\tt #1}}\fi
\expandafter\ifx\csname urlprefix\endcsname\relax\def\urlprefix{URL }\fi
\providecommand{\eprint}[2][]{\url{#2}}

\bibitem{meshik01}
Meshik A~P {\em et~al.\/} 2001 {\em Phys. Rev. \rm C\/} {\bf 64} 035205

\bibitem{pujol09}
Pujol M {\em et~al.\/} 2009 {\em Geochim. Cosmochim. Acta\/} {\bf 73}
  6834--6846

\bibitem{gavrilyuk13}
Gavrilyuk {\relax Yu}~M {\em et~al.\/} 2013 {\em Phys. Rev. \rm C\/} {\bf 87}
  035501

\bibitem{sujkowski03}
Sujkowski Z and Wycech S 2004 {\em Phys. Rev. \rm C\/} {\bf 70} 052501

\bibitem{krivoruchenko10}
Krivoruchenko M~I {\em et~al.\/} 2011 {\em Nucl. Phys. \rm A\/} {\bf 859}
  140--171

\bibitem{fang12}
Fang D~L {\em et~al.\/} 2012 {\em Phys. Rev. \rm C\/} {\bf 85} 035503

\bibitem{suhonen12}
Suhonen J 2012 {\em Eur. Phys. J. \rm A\/} {\bf 48} 51

\bibitem{rodriguez12}
Rodriguez T~R and Martinez-Pinedo G 2012 {\em Phys. Rev. \rm C\/} {\bf 85}
  044310

\bibitem{maalampi13}
Maalampi J and Suhonen J 2013 {\em Adv. High Energy Phys.\/} {\bf 2013} 505874

\bibitem{barabash06}
Barabash A~S {\em et~al.\/} 2007 {\em Nucl. Phys. \rm A\/} {\bf 785} 371--380

\bibitem{barabash08}
Barabash A~S {\em et~al.\/} 2008 {\em Nucl. Phys. \rm A\/} {\bf 807} 269--281

\bibitem{barabash09}
Barabash A~S {\em et~al.\/} 2009 {\em Phys. Rev. \rm C\/} {\bf 80} 035501

\bibitem{belli11}
Belli P {\em et~al.\/} 2011 {\em J. Phys. {\rm G}: Nucl. Part. Phys.\/} {\bf
  38} 115107

\bibitem{belli13}
Belli P {\em et~al.\/} 2013 {\em Phys. Rev. \rm C\/} {\bf 87} 034607

\bibitem{belli14}
Belli P {\em et~al.\/} 2014 {\em Nucl. Phys. \rm A\/} {\bf 930} 195--208

\bibitem{kotila14}
Kotila J {\em et~al.\/} 2014 {\em Phys. Rev. \rm C\/} {\bf 89} 064319

\bibitem{capturat}
Kantele J 1995 {\em {Handbook of Nuclear Spectrometry}\/} (Academic Press)

\bibitem{berglund11}
Berglund M and Wieser M~E 2011 {\em Pure Appl. Chem.\/} {\bf 83} 397--410

\bibitem{wang12}
Wang M {\em et~al.\/} 2012 {\em Chin. Phys. \rm C\/} {\bf 36} 1603

\bibitem{belli99}
Belli P {\em et~al.\/} 1999 {\em Nucl. Phys. \rm B\/} {\bf 563} 97--106

\bibitem{xrdbl}
Thompson A~C and Vaughan D (eds) 2001 {\em {X-ray Data Booklet}\/} 2nd ed
  (Lawrence Berkeley National Laboratory, University of California)

\bibitem{ching84}
Cheng-rui C {\em et~al.\/} 1984 {\em Comm. Theor. Phys.\/} {\bf 3} 517--520

\bibitem{droese11}
Droese C {\em et~al.\/} 2012 {\em Nucl. Phys. \rm A\/} {\bf 875} 1--7

\bibitem{iachello}
Iachello F private communication

\bibitem{angloher15}
Angloher G {\em et~al.\/} (CRESST) 2016 {\em Eur. Phys. J. \rm C\/} {\bf 76} 25

\bibitem{angloher09}
Angloher G {\em et~al.\/} (CRESST) 2009 {\em Astropart. Phys.\/} {\bf 31}
  270--276

\bibitem{angloher12}
Angloher G {\em et~al.\/} (CRESST) 2012 {\em Eur. Phys. J. \rm C\/} {\bf 72}
  1--22

\bibitem{nudat}
Chu S~Y~F {\em et~al.\/} {\em {WWW Table of Radioactive Isotopes, database
  version 1999-02-28 from URL
  http://nucleardata.nuclear.lu.se/nucleardata/toi/}\/}

\bibitem{strauss14}
Strauss R {\em et~al.\/} (CRESST) 2015 {\em JCAP\/} {\bf 1506} 030

\bibitem{agostinelli02}
Agostinelli S {\em et~al.\/} (GEANT4) 2003 {\em Nucl. Instrum. Meth. \rm A\/}
  {\bf 506} 250--303

\bibitem{ponkratenko00}
Ponkratenko O~A {\em et~al.\/} 2000 {\em Phys. Atom. Nucl.\/} {\bf 63}
  1282--1287

\bibitem{caldwell08}
Caldwell A {\em et~al.\/} 2009 {\em Comput. Phys. Commun.\/} {\bf 180}
  2197--2209

\bibitem{beaujean11}
Beaujean F {\em et~al.\/} 2011 {\em Phys. Rev. \rm D\/} {\bf 83} 012004

\bibitem{munster14}
M\"{u}nster A {\em et~al.\/} 2014 {\em JCAP\/} {\bf 1405} 018

\end{thebibliography}

\end{document}